\documentclass[twocolumn,preprintnumbers,amsmath,amssymb]{revtex4}
\usepackage{graphicx}
\usepackage{dcolumn}
\usepackage{bm}
\usepackage{pstricks}
\usepackage{epsfig}
\usepackage{psfrag}

\begin{document}

\preprint{Applied Physics Letters}

\title{Single shot measurement of a silicon single electron transistor}

\author{D. G. Hasko$^1$}
\email{dgh4@cam.ac.uk}
\author{T. Ferrus$^2$}
\email{taf25@cam.ac.uk}
\author{Q. R. Morrissey$^3$}
\author{S. R. Burge$^3$}
\author{E. J. Freeman$^3$}
\author{M. J. French$^3$}
\author{A. Lam$^1$}
\author{L. Creswell$^1$}
\author{R. J. Collier$^1$}
\author{D. A. Williams$^2$}
\author{G. A. D. Briggs$^4$}
\affiliation{$^1$Cavendish Laboratory, University of Cambridge, J. J. Thomson Avenue, CB3 0HE, Cambridge, United Kingdom}
\affiliation{$^2$Hitachi Cambridge Laboratory, J. J. Thomson Avenue, CB3 0HE, Cambridge, United Kingdom}
\affiliation{$^3$Rutherford Appleton Laboratory, Chilton, Didcot, OX11 0QX, Oxon, United Kingdom}
\affiliation{$^4$Department of Materials, University of Oxford, Parks Road, Oxford OX1 3PH, United Kingdom}

\date{\today}

\begin{abstract}

We have fabricated a custom cryogenic Complementary Metal-Oxide-Semiconductor (CMOS) integrated circuit that has a higher measurement bandwidth compared with conventional room temperature electronics. This allowed implementing single shot operations and observe the real-time evolution of the current of a phosphorous-doped silicon single electron transistor that was irradiated with a microwave pulse. Relaxation times up to 90\,$\mu$s are observed, suggesting the presence of well isolated electron excitations within the device. It is expected that these are associated with long decoherence time and the device may be suitable for quantum information processing.

\end{abstract}

                          
                            
\maketitle

The requirement for faster measurement usually results from the need to shorten the timescale of complex experiments or to access dissipative information in time-dependent measurements. However, the measurement bandwidth is often limited by the low pass filter effect resulting from the load capacitance due to the connecting cables and measurement circuit and the feedback resistor of the amplifier. This problem is acute for quantum devices operating in the pA range and especially in quantum information processing for which fast access to qubit states is essential. 

In solid-state quantum computation, single electron transistors (SET) \cite{Fulton} are now widely employed as a qubit readout device.\cite{SET}  However, the bandwidth of the current measurement of the SET is often smaller than the decoherence rate of the system.\cite{Zurek} This is due to its tunnel barrier being much larger than the quantum resistance $h/e^2$ and due to large parasitic resistances \cite{Beenakker}, but also by the necessity to operate the SET at low temperature. Thus, a direct measurement of the qubit state or its evolution in real time is not possible and it is usually monitored indirectly by measuring the time averaged SET current over a large number of identical qubit operations, each operation including initialization, manipulation and measurement stages.\cite{Gorman} Systematic changes in qubit operation, usually in the manipulation stage, are then used to characterize the behaviour of the qubit. Alternative methods based on radio frequency techniques \cite{RFSET} offer a much higher bandwidth, but can lead to unwanted changes in the quantum mechanical state.

The problem of such fast current measurements is addressed in this letter through the use of a custom CMOS integrated circuit based on charge integration that operates at 4.2\,K (LTCMOS) and in close proximity to a silicon SET. This arrangement enabled an efficient noise gain suppression by using shorter cabling between the measurement circuit and the device and a significant improvement in measurement speed ($\sim 10^4$ times faster \cite{Gorman}) compared to conventional measurements using room temperature source measure units. Quantum states in the SET were mapped by continuous microwave spectroscopy and the real time evolution of a particularly long lived state was observed in a single shot measurement.

The SETs are fabricated from a silicon-on-insulator (SOI) substrate with a 30 nm-thick silicon layer, doped with phosphorous at a density of $10^{20}$\,cm$^{-3}$. High resolution electron beam lithography and reactive ion etching were used to pattern a single dot of diameter $\sim$60\,nm. The 30-nm width constrictions that separate the dot from the connecting source and drain leads are depleted of electrons and act as tunnel barriers. Thermal oxidation is then used to reduce the dot size and tunnel barrier widths as well as for surface passivation, resulting in a significant reduction in random telegraph switching. The devices are controlled by an in-plane gate that is formed from the same SOI layer (Fig.\,1a and b). Devices were mounted in 20 pin leadless header packages at the end of the custom probe. The microwave signal was coupled to the SET by using an open ended semi-rigid waveguide that terminates $\sim$0.5\,mm away from the silicon surface (Fig. 1c). A LTCMOS circuit was incorporated into the probe and was used both to provide the voltages for the source, drain and gate leads and to measure the SET current. All lines were filtered by single stage low-pass filter with a cut-off of about 2\,MHz to suppress electrical noise from the CMOS circuit. Both the device and the LTCMOS were kept at 4.2\,K by immersing the probe into liquid helium.

\begin{figure}
\begin{center}
\includegraphics[width=86mm]{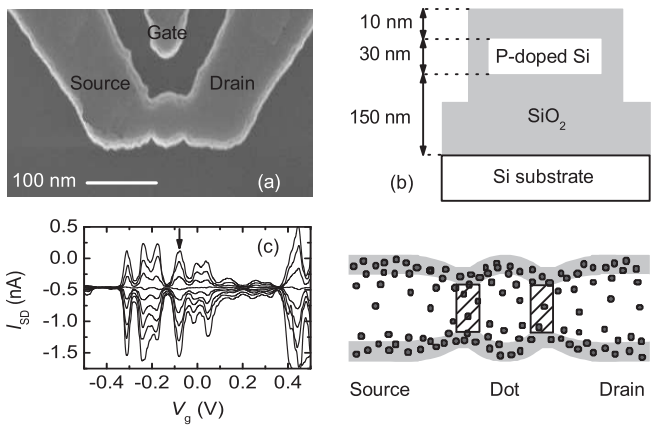}
\end{center}
\caption{\label{fig:figure1} (a) Scanning electron microscopy image of the SET. (b) Cross-section and plane view of the device with the sidewall depletion (grey), the source and drain potential barriers (dashed) and localising centres (black dots). (c) Source-drain current $I_{\textup{SD}}$ versus gate voltage $V_\textup{g}$ for source-drain biases $V_{\textup{SD}}$ from -2\,mV (bottom) to 2\,mV (top). The arrow indicates the SET working point.}
\end{figure}

The SET response to continuous microwave (CW) excitation consists of a large number of resonances that are detectable when the microwave power at the source $P$ is greater than about -20\,dBm (Fig. 2).\cite{Creswell} These resonances exhibit a wide range of amplitudes and quality factors $Q$ with the largest amplitudes corresponding to the lowest $Q$ values. Some low $Q$ resonances, typically below $10^4$, could be explained by cavity resonances, as these influence the frequency dependence of the coupled microwave power responsible for electron heating in the SET island.\cite{Manscher} A rise in electron temperature would effectively lead to an increase in the conductance of the SET through cotunneling and sequential tunneling effects.\cite{Averin} However this mechanism is inconsistent with the observed experimental behaviour in which resonances can be either an increase or a decrease in the SET current. Furthermore, in a few cases, the SET current at the resonance is even seen to reverse with respect to the source-drain potential difference. Such a behaviour suggests that at least some of these low $Q$ resonances are due to current pumping.

\begin{figure}
\begin{center}
\includegraphics[width=86mm]{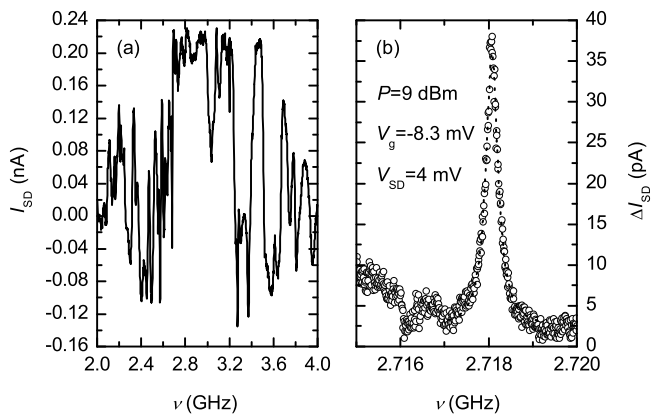}
\end{center}
\caption{\label{fig:figure2} (a) $I_{\textup{SD}}$ dependence on microwave frequency $\nu$, measured relatively to its value for $P=-20$\,dBm. (b) High $Q$ resonance used for the single shot measurement ($Q\sim41,000$). The Lorentzian fit is shown by the dotted line.}
\end{figure}

However the presence of resonances with large $Q$ values suggests a mechanism that is well isolated from the electromagnetic environment such as tunneling or trap-mediated transport. Such processes are indeed favoured by the presence of phosphorous donors in the SET island and tunneling barriers and by a significant number of localised states especially on the sidewalls of the SET.\cite{defect} This effect is enhanced by the large surface state density found on the non-(100) surfaces forming the sidewalls. Also, the resonance shapes are found to be Lorentzian with a width giving an energy scale two or three orders of magnitude lower than $k_{\textup{B}}T$. This is inconsistent with phonon-mediated hopping. However, according to Landauer theory, the conductance of a single channel wire with a barrier is given by the transmission probability \cite{Landauer}, and, for a sufficiently high barrier relatively to the Fermi energy, this probability is energy-independent, giving a temperature independent width in the resonance spectrum.\cite{barrier} The previous observation also implies a weak coupling to the phonon bath and non-adiabatic processes (e.g. the timescale for the transport mechanism is shorter than the thermalization time). In this case, the resonance widths $\Delta\nu_{\textup{0}}$ are related to the energy loss mechanism in the system and $Q=\pi \nu_{\textup{0}}/\Delta\nu_{\textup{0}}\,\sim\,\pi T_{\textup{0}}\nu_{\textup{0}}$, where $T_{\textup{0}}$ is the energy relaxation time of the system \cite{lifetime} and $\nu_{\textup{0}}$ the resonance frequency. Of the plausible transport mechanisms, trap-assisted tunneling in the tunnel barrier or potential shifts due to reversible dipoles \cite{Muller} in the SET island are most likely to be present as we did not observe any photon-assisted tunneling effects in the gate dependence of the source-drain current.\cite{PAT} Electron hops most likely occur between localised states near the SET island. This situation and the dependence of $\nu_0$ with gate voltage has been previously discussed.\cite{Creswell} The electron displacement modifies the electrostatic environment inside the SET island and leads to a change in the potential close to the tunnel barrier. This changes the magnitude of the barrier and is responsible for the increase or decrease in the measured current depending on the location of the active traps. Localised states in the direct transport pathway would lead to low $Q$ resonances with large amplitude due to the short residence time of the electron, whereas remote traps could be responsible for the high $Q$ but low amplitude resonances.

\begin{figure}
\begin{center}
\includegraphics[width=86mm]{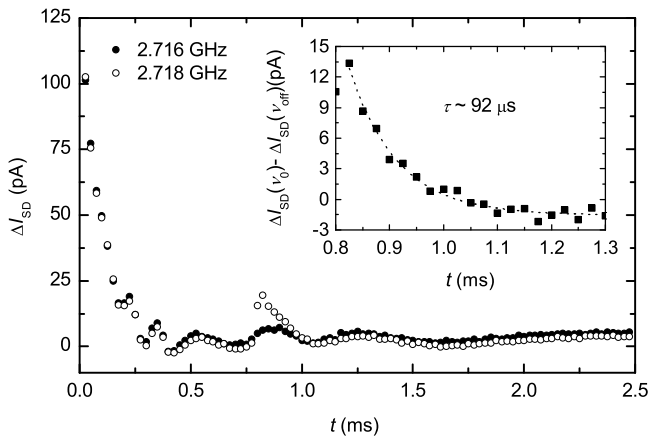}
\end{center}
\caption{\label{fig:figure3} Relative variation of $I_{\textup{SD}}$ with time $t$ for $\nu_{\textup{off}}=2.716\,$GHz (off resonance) and $\nu_{\textup{0}}=2.718\,$GHz (on resonance). The inset shows the exponential decay in the current.}
\end{figure}

The quality factor for the high $Q$ peaks suggests that the lifetime for these excitations can be as long as 100\,$\mu$s. This allows the study of the real time evolution of the source-drain current after the application of a short microwave pulse, known as a single shot measurement \cite{singleshot}. To this purpose, a well-isolated resonance on a level background at a frequency $\nu_{\textup{0}}\sim2.718$\,GHz was selected from a CW measurement at a power of +9\,dBm, with a charge integration time of 200\,$\mu$s and averaging the results over 200 measurements (Fig. 2). This results in a signal-to-noise ratio of about 100:1. The width of the chosen resonance was about 207\,kHz, for which the corresponding relaxation time is $T_{\textup{0}}\sim5\,\mu$s. A single shot measurement was then performed by reducing the charge integration time to 20\,$\mu$s and using a microwave pulse of duration of 2\,$\mu$s. The microwave power was increased to about +25\,dBm using a power amplifier to compensate for the decrease in the signal amplitude. These conditions also greatly reduced the signal to noise ratio compared to the CW measurement, so that the averaging had to be increased to 1000 measurements per point. The delay in the trigger signal causes the microwave pulse to arrive at $t\sim0.8\,$ms (Fig. 3). The signal measured in single shot mode contains some very large damped oscillations, which do not depend on the nature of the microwave pulse. They are even present in the absence of the microwave pulse and are measurement artefacts due to cable capacitance charging effects. By tuning the microwave frequency to on or off resonance values the effects of these background oscillations may be removed. The difference between these is an exponential decay with a time constant $\tau\sim90\,\mu$s. This value is more than one order of magnitude larger than that found from the CW linewidth measurement. As the time-averaged power in the pulsed measurement is a factor of 1000 less than that used in the CW measurement, this difference in relaxation time may result from microwave-induced electron heating in the CW measurement. The lifetime $\tau$ corresponds to the time for the microwave-excited electron to recombine with its original ionised trap. Its large value suggests glassy behaviour \cite{glass}, a characteristic manifestation of Coulomb interaction in systems with localized states. The disorder is due to the random placement of phosphorous donors and the glassy dynamics would result from the competition between the disorder and the Coulomb interaction.\\
In conclusion, we have been able to observe the real-time decay of a quantum state by performing a single shot measurement in a single electron transistor and using a low temperature custom CMOS measurement circuit. The microwave excitation induces electron hops between localised states within the SET island. The long relaxation time we observed suggests the presence of coherent transport in the structure, making the present system usable quantum information technology.\\
The authors thank EPSRC (UK) for funding under grant numbers GR/S24275/01 and GR/S15808/01. This work was partly supported by Special Coordination Funds for Promoting Science and Technology in Japan. Two of the authors (D. G. H. and T. F.) contributed equally to this work.

\end{document}